\documentclass[a4paper, 11pt]{article}

\title{
Contest success functions with(out) headstarts\thanks{The author thanks Anna Bogomolnaia, Hervé Moulin, Constantine Sorokin, Georgios Gerasimou, Takashi Hayashi, Sean Horan (Editor), and all anonymous referees for helpful suggestions. All remaining errors are my own.}
}

\author{
Hao Yu\thanks{Adam Smith Business School, University of Glasgow. Email: \href{mailto:hao.yu@glasgow.ac.uk}{hao.yu@glasgow.ac.uk}}
}

\date{\today}

\usepackage[margin=1.25in]{geometry}
\selectfont
\usepackage{amsmath, amsfonts, amsthm, amssymb, graphicx, enumitem}
\usepackage[colorlinks=true, citecolor=blue]{hyperref}
\usepackage[round]{natbib}

\usepackage{libertine}

\newtheorem{definition}{Definition}
\newtheorem{theorem}{Theorem}
\newtheorem{proposition}{Proposition}
\newtheorem{corollary}{Corollary}
\newtheorem{lemma}{Lemma}
\usepackage{lipsum}

\begin{document}

\maketitle
\begin{abstract}
     Contest success function (CSF) maps contestants' efforts to their winning probability. This paper provides axiomatizations of CSFs with headstarts. The results extend the classic axiomatization of the Tullock CSF and connect to CSFs that allow for draws. The central axiom is relative homogeneity of counterfactual deviation, which requires the pairwise influence of one contestant's effort on opponent's probabilistic allocation to be scale-invariant. Two fairness axioms and no advantageous reallocation further restrict the admissible functional forms with headstarts. We also introduce dummy consistency, requiring allocations to be consistent with and without inactive contestants, to clarify the relationship with earlier axiomatic work that rules out headstarts. Finally, we discuss an extension that drops the assumption of full allocation. \\[0.5cm]
\end{abstract}

\newpage

\section{Introduction}
Contest is a mechanism for compensation and rewards in which contestants exert irreversible effort for a chance to obtain a prize. It captures conflicts of interest among competing parties in many real-world settings.
In rent-seeking, firms offer bribes to increase their probability of securing a fixed policy rent. In patent races, firms invest in research and development to increase their chance of being first. In job promotion, workers exert effort to increase their probability of ranking first and being promoted.
These examples differ in context but share the same idea: agents compete for an object, and outcomes depend on their relative performance.

The mapping from contestants' efforts to winning probabilities is called contest success function (CSF). 
It is a reduced-form object that captures how interactions among agents shape a probabilistic allocation. Many widely used specifications fall into the asymmetric logit class,
$p^N_i(x)= f_i(x_i)\big/\sum_{j\in N} f_j(x_j)$,
where $p^N_i(x)$ is contestant $i$'s winning probability given the effort profile $x$ and $f_i(\cdot)$ is contestant $i$'s impact function. For example, \cite{tullock1980}'s $f_i(x_i)=x_i^r$, where $r$ is a discrimination parameter, is widely used in the early contest literature.
As a tractable alternative, \cite{amegashie2006} studies $f_i(x_i)= b + x_i$ for a constant noise/luck $b$, which allows a contestant to be selected no matter how small the effort is. The common baseline $b$ provides an early prototype of headstarts.

A large applied literature models favoritism, affirmative action, and related biases by postulating biased CSFs in which contestants receive multiplicative weights and/or additive headstarts \citep{chowdhury2023}.
However, headstarts are typically not effort-maximizing, because they partially substitute for contestants' investment in effort \citep{franke2018}. Related results supporting the optimality of non-headstart designs appear also in \cite{dasgupta1998} and \cite{nti2004}. 
\cite{kirkegaard2023} emphasizes that, while stochastic performance can microfound the unbiased lottery CSF, comparable foundations for biased lottery CSFs with headstarts are largely absent, leaving the approach open to the criticism that it is ad hoc. Moreover, because actions are typically unobservable and prizes are awarded based on noisy performance signals, interpreting ``biases'' or ``headstarts'' as direct adjustments to effort can be conceptually delicate.

This paper provides an axiomatic foundation for a broad class of biased lottery CSFs with headstarts. Rather than taking ticket counts as primitives, we impose invariance restrictions directly on counterfactual deviations that measure how one contestant's activity affects others' allocations. Our main theorem delivers the additive headstart structure $f_i(x_i)=b_i+a_i x_i^r$ with constants $r$, $a_i$, and $b_i$, thereby rationalizing the canonical biased lottery form used in applications. In particular, choosing the discrimination parameter $r$, bias parameters $a_i$, and headstart parameters $b_i$, all encompassed by our main CSF, are common approaches to leveling the playing field \citep{chowdhury2023}.

From the contest sponsor's perspective, the contest can also be interpreted as a random choice of a winner (alternative). Our CSFs can be viewed as natural extensions of conventional Luce/logit choice models in which each alternative has a continuous ``effort'' feature.
To our knowledge, the closest related framework is \cite{kovach2022}'s focal Luce model.
They propose a menu-dependent random choice model capturing decision-maker's biased focality over the choice set. Aligning with their model, the headstart $b_i$ in our framework captures the sponsor's baseline utility for choosing a certain winner, while contestants' efforts have heterogeneous ability to influence the sponsor's focal bias.\footnote{For example, we can rewrite our impact function as $f_i(x_i) = b_i[1+(a_i/b_i)x_i^r]$ if $b_i>0$ for all $i\in N$. Taking binary efforts $x$, the focal set $A$ is $\{i\in N: x_i>0\}$, $b_i$ corresponds to unbiased term $u(i)$, and $a_i/b_i$ corresponds to the focal bias $\delta (A)$ in \cite{kovach2022}. See \cite{hagiwara2024} for a status quo extension. We only highlight the similarity between our function to theirs, while the axiomatizations in two models are quite different. Zero probability of the choice is also considered in our CSFs. For other general Luce models handling zero probabilities, see e.g. \cite{ahumada2018}, \cite{echenique2019}, and \cite{horan2021}.
See also \cite{declippel2024} for a review of bounded rationality in choice theory, and \cite{strzalecki2025} for a review of random choice theory.
} 
The biased part caused by attractive efforts, however, meets the rewarding purpose of the contest. The CSFs with headstarts axiomatized here also admit a two-stage interpretation: the sponsor first chooses within a focal set and then chooses from the full set if no alternative is selected at the first stage.

Formal axiomatic works on CSFs begin with \cite{skaperdas1996}, which characterizes the symmetric logit, Tullock, and logistic CSFs. \cite{clark1998} extend this approach by dropping anonymity (ANY) and axiomatize the asymmetric Tullock CSF. Relaxing the assumption of full allocation, \cite{blavatskyy2010} provides an axiomatization of CSFs that allow for draws.
Building on these seminal contributions, this paper proposes alternative axioms related to homogeneity (HOM), as well as ANY, and generalizes a conventional class of CSFs, thereby enriching the axiomatic contest literature.

The key axiom is relative homogeneity of counterfactual deviation (d-RH). The counterfactual deviation $d_{ij}(x)$ measures the proportional change in contestant $j$'s allocation relative to the counterfactual in which contestant $i$ is inactive, namely $[p^N_j(0, x_{-i})-p^N_j(x)]/p^N_j(0, x_{-i})$. It captures the effort-based power of contestant $i$ over contestant $j$. The axiom requires that this relative power be invariant to common rescaling of efforts.
Surprisingly, restricting effort-based power still leaves room for headstarts that are not generated by effort. Under the asymmetric logit CSF, d-RH requires the ratio $[f_i(x_i)-f_i(0)]/[f_j(x_j)-f_j(0)]$ to be homogeneous. By contrast, \cite{skaperdas1996}'s HOM and \cite{blavatskyy2010}'s relative homogeneity of winning probability (p-RH) require $f_i(x_i)/\sum_{j\in N} f_j(x_j)$ and $f_i(x_i)/f_j(x_j)$ to be homogeneous, respectively, which rule out any functional form with $f_i(0)>0$.

Symmetric counterfactual deviation (d-SYM), proposed as an alternative to ANY, captures a notion of secondary fairness: it requires the effort-based power from one contestant to another to be symmetric, while allowing for unequal initial rationing by the sponsor. We further impose no advantageous reallocation (NAR), which requires that reallocating total efforts between two contestants does not affect others' allocations. It yields a tractable functional form with headstarts.

To clarify the axiomatic difference between CSFs with and without headstarts, we introduce dummy consistency (DC). This axiom equates being inactive with not participating. It not only connects our framework to earlier axiomatic works in the contest literature \citep{skaperdas1996, clark1998}, but also highlights a key distinction from other rationing models, where DC is typically assumed \citep{moulin2002}.\footnote{See \cite{moulin1987} for a comprehensive axiomatic analysis of surplus sharing. DC being not restricted is also the reason we get headstarts from axiomatization through NAR.} 
We also connect headstarts to models with a null outcome, aligning with the ``draw'' or ``outside option'' extensions discussed in literature \citep{blavatskyy2010}.

The rest of the paper is organized as follows. Section \ref{sec:ind1} introduces the model, axioms, and the main theorem. Section \ref{sec:ind2} provides more axiomatization results of CSFs with headstarts. Section \ref{sec:ind3} makes a discussion with previous works, and connects all results through an introduced axiom. Section \ref{sec:ind4} generalizes the model by allowing partial allocation. Section \ref{sec:ind5} concludes the paper. 

\section{Axiomatization} \label{sec:ind1}
We consider a contest that allocates a single prize among $n\geq 3$ contestants in the set $N$. Each contestant $i\in N$ invests irreversible effort $x_i$, and the contest sponsor randomly selects a winner from $N$ with probabilities that depend on the effort vector $x\equiv (x_i)_{i\in N}$. We also consider sub-contests among $m\geq 2$ contestants $M\subseteq N$, interpreted as situations in which the sponsor does not observe the efforts of contestants in $N\setminus M$. We restrict attention to active contests, meaning that at least one contestant exerts positive effort. Denote by $X\equiv\mathbb{R}_+^N\setminus\{\mathbf{0}\}$ the domain of $x$,\footnote{Domain of contestants' efforts $x$ is commonly implicitly given in literature. The focus of axiomatization is on continuous functional forms, so that allocation in inactive contest are avoided due to discontinuity at zero. The domain is $\mathbb{R}_{++}^N$ in Theorem 2 of \cite{skaperdas1996}, as Theorem 1, which helps the proof of Theorem 2, requires symmetric impact function to be positive.
The domain can be restricted into at least two positive efforts in \cite{clark1998} due to their axiom A1 requiring $p^N_i(x) < 1$, which means the situation of only one contestant being active and getting full allocation is not the case. \cite{blavatskyy2010} considers a different model with full domain, i.e. $\mathbb{R}_+^N$. Current work considers a nontrivial domain with at least one positive effort. 
Readers can check that the range of parameters would shrink for all functional forms with headstarts if extending to the full domain. 
For other continuous functional forms without headstarts in this paper, the largest domain is $X$. 
}
and by $x^M$ the projection of $x$ on $M\subseteq N$. 
A CSF maps an effort vector to winning probabilities. The axiomatization focuses on the continuous functional forms. The formal definition is shown as follows.

\begin{definition}
    Fix contestants $N$. A contest success function (CSF) is a collection of functions $p^M\equiv (p^M_i)_{i\in M}$ for all $M\subseteq N$, $|M|\geq 2$, such that $p^M_i: \mathbb{R}_+^M\setminus\{\mathbf{0}\}\rightarrow [0, 1]$ for all $i\in M,$ and $\sum_{i\in M} p^M_i = 1$.
\end{definition}

\subsection{Standard axioms}
Following axioms are adopted from \cite{skaperdas1996}:\footnote{
Note that the ``increasing'' used in axiom A2 of \cite{skaperdas1996} exactly means ``strictly increasing.'' Also, axioms A4 and A5 are expressed together as Luce's choice axiom shown in this paper. Note also that the proof technologies through Luce's choice axiom can only apply to the contest with $n\geq 3$ contestants. \cite{blavatskyy2010} provides counterexamples for both \cite{skaperdas1996} and \cite{clark1998}'s axiomatizations in terms of $n= 2$. We also provide a counterexample for our axiomatization in the case of $n=2$ in Appendix \ref{apx:ind1.4}.
}

\noindent\textbf{Strict monotonicity (SM)}: For all $x\in X$, all $i\in N$ such that $p^N_i(x) < 1$, and all $x_i'>0$, 
\begin{equation*}
    x_i < x_i'\quad \Rightarrow \quad p^N_i(x) < p^N_i(x_i', x_{-i}).
\end{equation*}

\noindent\textbf{Luce's choice axiom (LCA)}: For all $x\in X$, all $M\subseteq N$, and all $i\in M$, 
\begin{equation*}
    p^N_i(x) = p^M_i(x^M)p^N_M(x),
\end{equation*} 
where $p^N_M(x) \equiv \sum_{j\in M} p^N_j(x)$.

SM says that the allocation is strictly increasing in contestant's own effort until getting the entire prize. This is an axiom for non-triviality of efforts in contest. LCA says that the final allocation of winning probability can be decomposed through an allocation process that allocates first to a sub-group of contestants and then allocates among them. These standard axioms restrict the CSF into an asymmetric logit form. 
\begin{lemma} \label{lemma:logit}
    SM and LCA hold if and only if for all $i\in M\subseteq N$,
    \begin{equation}
        p^M_i(x^M) = \frac{f_i(x_i)}{\sum_{j\in M} f_j(x_j)}, \label{eq:ind11}
    \end{equation}
    for some strictly increasing function $f_j(x_j)\geq 0$ for all $j\in N$.
\end{lemma}

Proof is shown in Appendix \ref{apx:ind1.1}. The impact function $f_i(x_i)$ is usually interpreted as the merit of contestant $i$.

\subsection{Relative homogeneity of counterfactual deviation}
The key axiom characterizing the asymmetric Tullock CSF with headstarts is defined through counterfactual deviation. The counterfactual deviation is defined as
\begin{equation*}
    d_{ij}(x) = \frac{p^N_j(0, x_{-i})-p^N_j(x)}{p^N_j(0, x_{-i})}.
\end{equation*} 
Given effort vector $x$, $p^N_j(x)$ is contestant $j$'s final probabilistic allocation, and $p^N_j(0, x_{-i})$ is the counterfactual allocation of contestant $j$ when contestant $i$ is inactive. Thus $d_{ij}(x)$ represents counterfactual percentage change in contestant $j$'s allocation as contestant $i$ transitions from inactivity to their effort level under profile $x$. 

The axiom is expressed in a pairwise way:

\noindent\textbf{Relative homogeneity of counterfactual deviation (d-RH)}:\footnote{Note that the condition of this axiom is equivalent to $p^N_i(0, x_{-j}) > p^N_i(x) >0$ and $p^N_j(0, x_{-i}) > p^N_j(x) > 0$, which can be simplified into $x_i, x_j>0$ given SM and LCA. See Lemma \ref{lem:ind5} in Appendix \ref{apx:ind3.2} for details. Similarly, the condition of relative homogeneity introduced later can also be implied from $x_i, x_j>0$ given SM. } For all $x\in X$, all $i,j\in N$ such that $d_{ij}(x),d_{ji}(x)> 0$, and all $\lambda > 0$, 
\begin{equation*}
    \frac{d_{ij}(x)}{d_{ji}(x)} = \frac{d_{ij}(\lambda x)}{d_{ji}(\lambda x)}.
\end{equation*}

This condition enlarges the functional forms axiomatized by homogeneity axioms in ways we explain below. It imposes an invariance restriction on the ratio of counterfactual proportional changes between any two active contestants $i$ and $j$ as all contestants' efforts are scaled by the same factor. 

\subsection{Axiom comparison}
To clarify the content of d-RH, we compare it with two familiar axioms from the literature, i.e. homogeneity and relative homogeneity \citep{skaperdas1996, clark1998, blavatskyy2010}. Two axioms are defined as follows:

\noindent\textbf{Homogeneity (HOM)}: For all $x\in X$, all $i\in N$, and all $\lambda > 0$, 
\begin{equation*}
    p^N_i(x) = p^N_i(\lambda x).
\end{equation*}

\noindent\textbf{Relative homogeneity (p-RH)}:\footnote{To distinguish it from the central axiom d-RH, the abbreviation of relative homogeneity is modified with a prefix. It represents the relative homogeneity of winning [p]robability in contrast to that of counterfactual [d]eviation.} For all $x\in X$, all $i,j\in N$ such that $p^N_i(x), p^N_j(x) > 0$, and all $\lambda > 0$, 
\begin{equation*}
    \frac{p^N_i(x)}{p^N_j(x)} = \frac{p^N_i(\lambda x)}{p^N_j(\lambda x)}.
\end{equation*}

HOM means that the probabilistic allocation is scale-invariant. p-RH states that the relative allocation is scale-invariant. Obviously, HOM implies p-RH. Given HOM, for all $\lambda > 0$, there is
\begin{equation*}
    d_{ij}(x) = \frac{p^N_j(0, x_{-i})-p^N_j(x)}{p^N_j(0, x_{-i})} = \frac{p^N_j(0,\lambda  x_{-i})-p^N_j(\lambda x)}{p^N_j(0, \lambda x_{-i})} = d_{ij}(\lambda x).
\end{equation*}
Thus d-RH is also implied by HOM.

However, neither axiom accommodates CSFs with headstarts. Consider a contest with $n = 3$ and $x = (2, 1, 0)$. Consider the CSF (\ref{eq:ind11}) with impact function $f_j(x_j) = 1 + x_j$ for all $j\in N$, which is a simplest example of CSFs with headstarts, In the full contest, the winning probability of contestant $i$ is
\begin{equation*}
    p^N_i(x) = \frac{1 + x_i}{\sum_{j\in N} (1 + x_j)}.\label{eq:mainL}
\end{equation*}
HOM requires that, for all $\lambda > 0,$
\begin{equation*}
    \frac{1 + x_i}{\sum_{j\in N} (1 + x_j)} = \frac{1 + \lambda x_i}{\sum_{j\in N} (1 + \lambda x_j)}.
\end{equation*}
Take $\lambda = 2$. For $i = 1$, the left-hand side (LHS) $= 1/2$,
and the right-hand side (RHS) $= 5/9 \neq$ LHS.
Thus HOM is violated. Similarly, p-RH requires that, for all $\lambda > 0,$
\begin{equation*}
    \frac{1 + x_i}{1 + x_j} = \frac{1 + \lambda x_i}{1 + \lambda x_j}.
\end{equation*}
Let $i = 1$, $j = 2$, and $\lambda=2$. LHS $= 3/2\neq 5/3 =$
RHS. Therefore, p-RH is also violated.

Now taking SM and LCA as given, Lemma \ref{lemma:logit} provides an asymmetric logit form (\ref{eq:ind11}). Since HOM implies p-RH, we only compare p-RH and d-RH here. On the basis of logit form, p-RH implies that
\begin{equation}
    \frac{f_i(x_i)}{f_j(x_j)} = \frac{f_i(\lambda x_i)}{f_j(\lambda x_j)},\label{eq:impact1}
\end{equation}
holds for all $i,j\in N, x_i, x_j > 0$, that is, pairwise comparison of impact function is homogeneous. Notice that the deviation is $d_{ij}(x) = [f_i(x_i) - f_i(0)]/\sum_{k\in N}f_k(x_k)$ given general logit form. Thus d-RH implies that
\begin{equation}
    \frac{f_i(x_i) - f_i(0)}{f_j(x_j) - f_j(0)} = \frac{f_i(\lambda x_i) - f_i(0)}{f_j(\lambda x_j) - f_j(0)},\label{eq:impact2}
\end{equation}
holds for all $i,j\in N$ such that $x_i, x_j > 0$. Comparing between (\ref{eq:impact1}) and (\ref{eq:impact2}), both of p-RH and d-RH provide a scale-invariant relative power, which are different in the source of such power. p-RH provides the merit-based power of contestant $i$ through $f_i(x_i)$, while d-RH constructs a power through the incremental merit $f_i(x_i)-f_i(0)$ created by contestant $i$'s effort. Thus the new axiom d-RH, different from p-RH, provides the room for the existence of headstarts $f_i(0)$ for all $i\in N$.

\subsection{The main theorem}
\begin{theorem} \label{thm:main}
    SM, LCA, and d-RH hold if and only if for all $i\in M\subseteq N,$ 
    \begin{equation}
        p^M_i(x^M) = \frac{b_i + a_i x_i^r}{\sum_{j\in M} (b_j + a_j x_j^r)},\label{eq:mainluck}
    \end{equation}
    for some $r > 0$, $a_j > 0$, and $b_j\geq 0$ for all $j\in N$.
\end{theorem}

\begin{proof}[Sketch of proof]
(only if) part includes 5 steps. Step 1 applies Lemma \ref{lemma:logit} to get the asymmetric logit form (\ref{eq:ind11}). Step 2 constructs a series of two-level CSFs, including individual CSFs and a CSF allowing for a draw among all contestants. Step 3 considers the case of the draw CSF being zero and gets the form of asymmetric Tullock CSF with $\sum_{j\in N} b_j = 0$, followed by a similar proof of \cite{clark1998}. Step 4 considers the case of the draw CSF being positive, applies \cite{blavatskyy2010}'s corollary, and gets the CSF with $\sum_{j\in N} b_j > 0$. Step 5 concludes the forms derived from former two steps, and uses LCA to get the sub-contest representations. 
\end{proof}

Two proofs are provided in Appendix \ref{apx:ind1}. 
The main proof, of which a sketch is shown here, reveals that the headstart part indeed exhibits a second-level allocation of remaining prize among all contestants, when the remaining prize is not allocated to any contestant in the first level. 
Alternatively, the functional form can also be obtained directly from some constructions.

This CSF is in a broader form than before. For example, (a)symmetric Tullock CSF is a special case of this CSF by restricting headstart part $b_i = 0$ for all $i\in N$ \citep{skaperdas1996, clark1998}, and thus a series of more specific CSFs are also covered. CSF with linear contest technology is a special case of this CSF by restricting $r=1$, which is also widely used in headstarts literature \citep{nti2004, franke2018}. 

\section{More specifications with headstarts} \label{sec:ind2}

\subsection{Two fairness axioms} 
So far the main theorem axiomatizes the asymmetric Tullock CSF with headstarts. This section considers two fairness axioms, including conventional anonymity and new symmetric counterfactual deviation. 

\noindent\textbf{Anonymity (ANY)}: For all $x\in X$ and all permutation $\pi$ on $N$,
\begin{equation*}
    p_\pi^N(x) = p^N(x_\pi),
\end{equation*}
where subscript is to represent the permutation on identities. 

This conventional axiom is in the same spirit of Aristotle's equal treatment of equals. That is, contestants who invest the same should have equal probability winning the prize. Under ANY, the symmetric Tullock CSF with equal starts is immediately derived.

\begin{corollary} \label{corollary:symm}
    SM, LCA, d-RH, and ANY hold if and only if for all $i\in M\subseteq N$, 
    \begin{equation}
        p^M_i(x^M) = \frac{b + x_i^r}{\sum_{j\in M} (b + x_j^r)},\label{eq:symluck}
    \end{equation}
    for some $b\geq 0$ and $r > 0$.
\end{corollary}

Another fairness axiom in the form of counterfactual deviation is newly introduced.

\noindent\textbf{Symmetric counterfactual deviation (d-SYM)}: For all $x\in X$ and all $i,j\in N$,
\begin{equation*}
    x_i = x_j',\ \mathrm{and}\ x_j = x_i'\quad \Rightarrow\quad d_{ij}(x_i, x_j, x^{N\setminus\{i,j\}}) = d_{ji}(x_i', x_j', x^{N\setminus\{i,j\}}).
\end{equation*}

This axiom considers a weaker situation by restricting only effort-based power to be fair instead of the final merit. It states a restricted fairness by ignoring the unfair endowed rationing. In real world, balancing the education resources throughout the whole country is usually unable to give two equally hard-working students the same score due to differences in prior knowledge, wealth, intelligence, and so on. This principle acknowledges these gaps and ensures the secondary fairness among contestants.

Under an asymmetric logit form, i.e. SM and LCA are given, d-SYM restricts the effort-based power to be symmetric. There is a symmetric function $g(x_i) = f_i(x_i)-f_i(0)$ for all $i\in N$, capturing the symmetric incremental merit. Define $b_i = f_i(0)$ for all $i\in N$. Following lemma can be obtained immediately.
\begin{theorem}\label{thm:ind2}
    SM, LCA, and d-SYM hold if and only if for all $i\in M\subseteq N,$
    \begin{equation}
        p^M_i(x^M) = \frac{b_i + g(x_i)}{\sum_{j\in M} [b_j + g(x_j)]},
    \end{equation}
    for some $b_j\geq 0$ for all $j\in N$ and strictly increasing function $g(\cdot)$ such that $g(0)=0$.
\end{theorem}

Combining both Theorems \ref{thm:main} and \ref{thm:ind2}, following corollary axiomatizing symmetric Tullock CSF with headstarts can be obtained.
\begin{corollary}
    SM, LCA, d-RH, and d-SYM hold if and only if for all $i\in M\subseteq N$,
    \begin{equation}
        p^M_i(x^M) = \frac{b_i + x_i^r}{\sum_{j\in M} (b_j + x_j^r)},\label{eq:ind11.2}
    \end{equation}
    for some $r >0$ and $b_j\geq 0$ for all $j\in N$.
\end{corollary}

\subsection{No advantageous reallocation}
Next axiom is derived from \cite{moulin1985, moulin1987}, that is,

\noindent\textbf{No advantageous reallocation (NAR)}: For all $x\in X$, all $i,j\in N$, all $x_i', x_j'\geq 0$ such that $x_i+x_j = x_i' + x_j'$, and all $k\in N\setminus\{i,j\}$, 
\begin{equation*}
    p^N_k(x_i, x_j, x^{N\setminus\{i, j\}}) = p^N_k(x_i', x_j', x^{N\setminus\{i,j\}}).
\end{equation*}

This axiom says that redistributing efforts between two contestants would not influence the probabilistic allocation among others, and thus generates no advantage from reallocation. It is a stronger axiom than d-SYM by restricting the influence from each amount of efforts, instead of the influence from just exchanging efforts. 
It restricts the functional form as follows:
\begin{theorem} \label{thm:headonly}
    SM, LCA, and NAR hold if and only if for all $i\in M\subseteq N$,
    \begin{equation}
        p^M_i(x^M) = \frac{b_i + x_i}{\sum_{j\in M} (b_j + x_j)},\label{eq:ind11.4}
    \end{equation}
    for some $b_j \geq 0$ for all $j\in N$.
\end{theorem}
Proof is shown in Appendix \ref{appendix:headonly}. Combining with ANY, the CSF with impact function $f_i(x_i) = b + x_i$ for some $b\geq 0$ is also characterized. For positive results by applying symmetric and asymmetric versions of this CSF, see \cite{amegashie2006}, and \cite{yu2026}, respectively.

\section{Dummy consistency and non-headstart specifications} \label{sec:ind3}
Comparing to the asymmetric Tullock CSF axiomatized before \citep{clark1998}, it's obviously that the main difference is on the headstarts. To figure out the axiomatic reason of such difference, we suggest a new axiom as follows, which is reasonable and helps to build a bridge between current and former axiomatic results.

\noindent\textbf{Dummy consistency (DC)}: For all $x\in X$, all $i\in N$, all $j\in N\setminus\{i\}$,
\begin{equation*}
    x_i=0\quad\Rightarrow \quad p^N_j(x) = p_j^{N\setminus\{i\}}(x^{N\setminus\{i\}}).
\end{equation*}

It says that the allocation for any contestant is irrelevant to inactive contestants (dummies), and such contest is equivalent to the sub-contest ruling out them. Without this axiom, it's hard to state the equivalence between being active and participating in a contest. Obviously, adding DC makes the impact function in (\ref{eq:ind11}) satisfy $f_i(0) = 0$ for all $i\in N$.

\cite{clark1998}'s special axiom and theorem are as follows.

\noindent\textbf{Clark-Riis independence (CRI)}: For all $x\in X$ and all pair $i,j\in N$, 
\begin{equation*}
    p^N_i(0,x_{-j}) = \frac{p^N_i(x)}{1 - p^N_j(x)}.
\end{equation*}

\begin{lemma}[\citealp{clark1998}] \label{lem:ind4}
    SM, CRI, and HOM hold if and only if for all $i\in N$,
    \begin{equation}
        p^N_i(x) = \frac{a_i x_i^r}{\sum_{j\in N} a_j x_j^r}, 
    \end{equation}
    for some $r>0$ and $a_j > 0$ for all $j\in N$.
\end{lemma}

In \cite{clark1998}, only CSF for full contest is given. Obviously, corresponding sub-contest representation can be derived by adding LCA. Comparing it with the axiomatic characterization in Theorem \ref{thm:main}, the non-existence of headstarts potentially depends on two axioms, i.e. CRI and HOM.

First, CRI has a direct influence on probabilistic allocation of inactive contestants. Such allocation must be zero, since this axiom restricts that the allocations of others remain the same in both full contest and sub-contest without inactive contestants. 

\begin{proposition} \label{prop:dummy}
    Taking LCA as given, CRI is equivalent to DC.
\end{proposition}

Proof is shown in Appendix \ref{appendix:dummy}. This proposition reveals the equivalence between two axioms with full contest and sub-contest representations. 

Interestingly, this axiom eliminates a lot of potential CSFs characterized by d-RH. Under CRI, the counterfactual deviation of contestant $j$'s allocation due to contestant $i$ being active would be
\begin{equation*}
    d_{ij}(x) = \frac{p^N_j(0, x_{-i}) - p^N_j(x)}{p^N_j(0, x_{-i})} 
    = \frac{\frac{p^N_j(x)}{1-p^N_i(x)} - p^N_j(x)}{\frac{p^N_j(x)}{1-p^N_i(x)}}
    = p^N_i(x).
\end{equation*}
In this case, counterfactual deviation coincides the final allocation. In other words, contestant $i$'s effort-based power over others coincides with contestant $i$'s final allocation in the non-headstart case. It immediately gives the following proposition.

\begin{proposition} \label{prop:homo_rela}
    Taking CRI as given, d-RH and p-RH are equivalent; d-SYM and ANY are equivalent.
\end{proposition}


Second, HOM may also restrict the CSF to the Tullock form. Recall that Lemma \ref{lem:ind4} and Proposition \ref{prop:dummy} imply a characterization of Tullock CSF, i.e. SM, LCA, DC, and HOM. Following proposition reveals that it is not tight.

\begin{proposition} \label{prop:dummy2}
    SM, LCA, and HOM imply DC.
\end{proposition}

Proof is shown in Appendix \ref{apx:ind3.2}. Applying Proposition \ref{prop:dummy} further, LCA and DC imply CRI. Combining SM, CRI, and HOM, the asymmetric Tullock CSF for full contest is derived from Lemma \ref{lem:ind4}. Sub-contest representations can be easily derived from LCA. Thus the following corollary holds.
\begin{corollary} \label{corollary:tullock}
    SM, LCA, and HOM hold if and only if for all $i\in M\subseteq N,$
    \begin{equation}
        p^M_i(x^M) = \frac{a_i x_i^r}{\sum_{j\in M} a_j x_j^r},
    \end{equation}
    for some $r>0$ and $a_j > 0$ for all $j\in N$.
\end{corollary}

Adding ANY, it is exactly the axiomatization of symmetric Tullock CSF \citep{skaperdas1996}.

\section{A generalized model allowing partial allocation}\label{sec:ind4}
In this section, we study a variant of \cite{blavatskyy2010}'s framework that allows for events in which no contestant is selected as the unique winner.\footnote{Thanks for an anonymous referee suggesting that.} 
Similar situations arise in chess or e-sports where matches may end in a draw; in auctions with a reserve price where the object may remain unsold; or in award committees and procurement processes where multiple candidates can be declared co-winners.

Aligning with \cite{blavatskyy2010}'s setting, we consider only CSFs in a presentation of full contest. Unlike the benchmark model in the previous sections, we allow partial allocation, meaning that the prize need not be fully allocated to contestants, i.e. $\sum_{i\in N} p^N_i(x) \leq 1$.
Define the null CSF $p^N_{\mathrm{null}}(x) \equiv 1 - \sum_{i\in N} p^N_i(x) \geq 0$. It captures all other the undefined results of a contest.

\noindent\textbf{Independence of irrelevant efforts (IIE)}:\footnote{IIE is a variant of \cite{blavatskyy2010}'s Axiom 3a, of which an explanation is shown in Appendix \ref{apx:ind1.1}. 
} 
For all $i\in I\subset N$, there exists a function $g_i^I:\mathbb{R}_+^I\rightarrow[0, 1]$ such that for all $x\in X$,
\begin{equation*}
    p^N_i(x) = g_i^I(x^I)[1 - p^N_{N\setminus I}(x)].
\end{equation*}

IIE states that the probability of contestant $i$ being chosen among $N$ conditional on contestants $N\setminus I$ not chosen is irrelevant to the efforts of contestants $N\setminus I$.

\begin{theorem}\label{thm:ind4}
    In a generalized model, SM and IIE hold if and only if for all $i\in N$,
    \begin{equation}
        p^N_i(x) = \frac{f_i(x)}{z + \sum_{j\in N} f_j(x)},
    \end{equation}
    for some $z\geq 0$, and strictly increasing function $f_j(x_j)\geq 0$ for all $j\in N$.     
\end{theorem}
Proof is shown in Appendix \ref{apx:ind8}. Following corollary can be obtained through the same steps as theorems or corollaries from previous sections.

\begin{corollary}
    In addition to Theorem \ref{thm:ind4}, 
    \begin{enumerate}[label=(\alph*)] 
        \item if p-RH holds, $f_j(x_j) = a_j x_j^r$, for some $a_j>0$ and $r>0$;
        \item if d-RH holds, $f_j(x_j) = b_j + a_j x_j^r$, for some $a_j>0$, $b_j\geq 0$, and $r>0$;
        \item if d-SYM holds, $f_j(x_j) = b_j + g(x_j)$, for some $b_j\geq 0$ and strictly increasing function $g(\cdot)$ such that $g(0)=0$;
        \item if NAR holds, $f_j(x_j) = b_j + x_j$, for some $a_j>0$.
    \end{enumerate}
\end{corollary}

\section{Concluding remarks} \label{sec:ind5}
This paper axiomatizes a class of contest success functions with headstarts, building on the foundational works of (a)symmetric logit and Tullock CSFs \citep{skaperdas1996, clark1998}. The core axiom, relative homogeneity of counterfactual deviation, ensures the relative effort-based power being scale-invariant. It allows the presence of headstarts, while traditional homogeneity and relative homogeneity eliminate all headstarts by restricting final allocations directly, either in absolute or relative sense.

The presence of headstarts can also be interpreted as a two-level allocation process: effort determines partial outcomes, while headstart governs the remaining prize share. This interpretation emerges in the proof of the main theorem via \cite{blavatskyy2010}'s Corollary 2, where individual CSFs map to the effort part, and the draw CSF captures the remainder composed by headstarts. In addition, anonymity, symmetric counterfactual deviation, and no advantageous reallocation can help to restrict the parameter space.

Dummy consistency plays the central role of bridging axiomatic works of CSFs with and without headstarts. It clarifies the difference between being active and participating in contest. It adopts the axiomatization of asymmetric Tullock CSF into a unified and discussable axiomatic structure. 

Further, we consider a generalized model by dropping the assumption of full allocation \citep{blavatskyy2010}. By substituting an independence axiom, we obtain similar but more general functional forms, providing a room for undefined events out of individual success, as those axiomatized under standard settings.

\bibliography{main}

\newpage
\appendix
\section{Proofs}
\subsection{Proof of Theorem \ref{thm:main}} \label{apx:ind1}
\subsubsection{Three lemmata}\label{apx:ind1.1}
Before formally expressing the proof of the main theorem, three lemmata are needed. Lemma \ref{lemma:logit} extends the traditional axiomatization of symmetric logit CSF provided by \cite{skaperdas1996} into the asymmetric one.

\begin{proof}[Proof of Lemma \ref{lemma:logit}]
(only if) Without loss of generality, let contestant 1 be always active, and take contestant 1's effort $\overline{x}_1>0$ as fixed. Consider arbitrary sub-contest $M\subseteq N$ that contains contestant 1. From SM, there is $p^N_1(x) > p^N_1(0, x_{-1}) \geq 0$, and thus $p^N_M(x) = p^N_1(x) + \sum_{i\in M\setminus\{1\}} p^N_i(x) > 0$. From LCA, there is $p_1^M(x^M) = p^N_1(x)/p^N_M(x) > 0$. 

Let $M = \{1, i\}$. LCA implies that
\begin{equation*}
    \frac{p^N_i(x)}{p^N_1(x)} 
    = \frac{p_i^{\{1, i\}}(x_i; \overline{x}_1)p_{\{1, i\}}^N(x)}{p_1^{\{1, i\}}(x_i; \overline{x}_1)p_{\{1, i\}}^N (x)} 
    = \frac{p_i^{\{1, i\}}(x_i; \overline{x}_1)}{p_1^{\{1, i\}}(x_i; \overline{x}_1)} 
    = \frac{p_i^{\{1, i\}}(x_i; \overline{x}_1)}{1 - p_i^{\{1, i\}}(x_i; \overline{x}_1)}.
\end{equation*}
Notice here that $i$ is arbitrarily chosen among $N\setminus \{1\}$. Normalize $f_1(\overline{x}_1) = 1$ and let $f_i(x_i)\equiv p^{\{1, i\}}_i(x_i; \overline{x}_1)/[1 - p^{\{1, i\}}_i(x_i; \overline{x}_1)]$ for all $i\in N\setminus \{1\}$. It implies that for all $i\in N$,
\begin{equation*}
    p^N_i(x) = p^N_1(x)f_i(x_i).
\end{equation*}
Notice that  $1 = \sum_{i\in N} p^N_i(x) = p^N_1(x) \sum_{i\in N} f_i(x_i)$. Thus for all $i\in N$,
\begin{equation}
    p^N_i(x) = \frac{f_i(x_i)}{\sum_{j\in N} f_j(x_j)}. \label{eq:logit}
\end{equation}

Now prove that $f_i(\cdot)$ is non-negative and strictly increasing. Derived from (\ref{eq:logit}), $f_i(x_i) = p^N_i(x) \sum_{j\in N} f_j(x_j)$, where $p^N_i(x) \geq 0$ and $\sum_{j\in N} f_j(x_j) > 0$. Thus $f_i(x_i) \geq 0$. In addition, if $x_i > 0$, there is $p^N_i(x) > 0$ and thus $f_i(x_i) = p^N_i(x) \sum_{j\in N} f_j(x_j) > 0$. Suppose there exists a point such that $x_i < x_i'$ and $f_i(x_i) \geq f_i(x_i')$. It implies that 
\begin{equation*}
    p^N_i(x) = \frac{f_i(x_i)}{f_i(x_i) + \sum_{j\in N\setminus\{i\}} f_j(x_j)} \geq \frac{f_i(x_i')}{f_i(x_i') + \sum_{j\in N\setminus\{i\}} f_j(x_j)} = p^N_i(x_i', x_{-i}).
\end{equation*}
Notice also that the axiom applies for all effort vectors with at least one active contestant, which contains $\sum_{j\in N\setminus\{i\}} f_j(x_j) > 0$ making $p^N_i(x) < 1$. So SM is violated. Therefore, the function $f_j(\cdot)$ is strictly increasing.

Finally, LCA is used to derive the sub-contest CSFs. That is, for all $i\in M\subseteq N$,
\begin{equation*}
    p^M_i(x^M) = \frac{p^N_i(x)}{p^N_M(x)} = \frac{ \frac{f_i(x_i)}{\sum_{k\in N} f_k(x_k)}}{\sum_{j\in M} \frac{f_j(x_j)}{\sum_{k\in N} f_k(x_k)}} = \frac{f_i(x_i)}{\sum_{j\in M} f_j(x_j)}.
\end{equation*}

(if) First check SM. For arbitrarily chosen contestant $i$, the effort $x_i$ must be positive when $\sum_{j\in N\setminus\{i\}} f_j(x_j) = 0$ in order to be feasible in domain, because $\sum_{j\in N\setminus\{i\}} f_j(x_j) = 0$ implies $x_j = 0$ for all $j\in N\setminus\{i\}$, and thus violates the assumption of at least one active contestant. In this case, the CSF of contestant $i$ will always be 1 and thus SM becomes trivial. If there exists one $x_j>0$ for some $j\neq i$, then for $x_i < x_i'$, there is $f_i(x_i) < f_i(x_i')$. Thus 
\begin{equation*}
    p^N_i(x_i, x_{-i}) = \frac{f_i(x_i)}{f_i(x_i) + \sum_{j\in N\setminus\{i\}} f_j(x_j)} < \frac{f_i(x_i')}{f_i(x_i') + \sum_{j\in N\setminus\{i\}} f_j(x_j)} = p^N_i(x_i', x_{-i}).
\end{equation*}
So SM is satisfied. Next, 
\begin{equation*}
    p^N_i(x) = \frac{f_i(x_i)}{\sum_{j\in N} f_j(x_j)} = \frac{f_i(x_i)}{\sum_{j\in M} f_j(x_j)} \sum_{i\in M} \frac{f_i(x_i)}{\sum_{j\in N} f_j(x_j)} = p^M_i(x^M) p^N_M(x).
\end{equation*}
Thus LCA is obtained.
\end{proof}

Second lemma required is an axiomatization of asymmetric Tullock CSF. It differs from \cite{clark1998}'s theorem in that CRI and HOM are replaced by LCA and p-RH, respectively. Also, it is expressed in a sub-contest representation.
\begin{lemma} \label{lemma:tullock}
    SM, LCA, and p-RH hold if and only if for all $i\in M\subseteq N,$
    \begin{equation}
        p^M_i(x^M) = \frac{a_i x_i^r}{\sum_{j\in M} a_j x_j^r},\label{eq:ind11.1}
    \end{equation}
    for some $r> 0$ and $a_j > 0$ for all $j\in N$.
\end{lemma}

\begin{proof}
(only if) From Lemma \ref{lemma:logit}, asymmetric logit CSF is derived as (\ref{eq:logit}) through SM and LCA, where the function $f_j(\cdot)$ is non-negative and strictly increasing. p-RH implies that
\begin{equation*}
    \frac{\frac{f_i(x_i)}{\sum_{k\in N} f_k(x_k)}}{\frac{f_j(x_j)}{\sum_{k\in N} f_k(x_k)}} = \frac{\frac{f_i(\lambda x_i)}{\sum_{k\in N} f_k(\lambda x_k)}}{\frac{f_j(\lambda x_j)}{\sum_{k\in N} f_k(\lambda x_k)}},
\end{equation*}
which is equivalent to 
\begin{equation}
    \frac{f_i(\lambda x_i)}{f_i(x_i)} = \frac{f_j(\lambda x_j)}{f_j(x_j)}, \label{eq:homo}
\end{equation}
for all $i,j\in N$ and $x_i, x_j > 0$. Notice that $x_i > 0$ is arbitrarily chosen in the LHS. There is
\begin{equation*}
    \frac{f_i(\lambda x_i)}{f_i(x_i)} = \frac{f_j(\lambda x_j)}{f_j(x_j)} = \frac{f_i(\lambda)}{f_i(1)},
\end{equation*}
where $f_i(1) > 0$ due to SM. Let $x_i = z$, it's equivalent to
\begin{equation*}
    \frac{f_i(\lambda z)}{f_i(1)} = \frac{f_i(z)}{f_i(1)} \frac{f_i(\lambda)}{f_i(1)}.
\end{equation*}
Let $F_i(z)\equiv f_i(z)/f_i(1)$, there is $F_i(\lambda z) = F_i(\lambda)F_i(z)$, which is one of fundamental Cauchy equations. Notice also $F_i(z)$ is strictly increasing due to the strictly increasing function $f_i(z)$. It turns out that the unique continuous form is $F_i(z) = z^{r_i}$ \citep{aczel1966}. Let $a_i = f_i(1) > 0$, then there is $f_i(x_i) = f_i(1)F_i(x_i) = a_i x_i^{r_i}$. Substituting it into (\ref{eq:homo}), it is
\begin{equation*}
    \frac{a_i (\lambda x_i)^{r_i}}{a_i x_i^{r_i}} = \frac{a_j (\lambda x_j)^{r_j}}{a_j x_j^{r_j}},
\end{equation*}
which implies $\lambda^{r_i} = \lambda^{r_j}$ for all $i, j\in N$ and $\lambda > 0$. Thus $r_i = r_j \equiv r$ for all $i, j\in N$. Substituting $f_i(x_i) = a_i x_i^r$ for all $i\in N$ into (\ref{eq:ind11}) ends the proof.

(if) Since it is a sub-class of (\ref{eq:ind11}), SM and LCA are satisfied from Lemma \ref{lemma:logit}. Now check p-RH. For arbitrarily chosen $i\in N$ and positive efforts $x_i> 0$, there is 
\begin{equation*}
    p^N_i(x) = \frac{a_i x_i^r}{\sum_{k\in N} a_k x_k^r} = \frac{\lambda^r a_i x_i^r}{\lambda^r \sum_{k\in N} a_k x_k^r} = \frac{a_i (\lambda x_i)^r}{\sum_{k\in N} a_k (\lambda x_k)^r}= p^N_i(\lambda x),
\end{equation*}
for any $\lambda > 0$. Arbitrarily choosing another $j\in N\setminus \{i\}$ with $x_j > 0$, there is 
\begin{equation*}
    \frac{p^N_i(x)}{p^N_j(x)} = \frac{p^N_i(\lambda x)}{p^N_j(\lambda x)},
\end{equation*}
and thus the p-RH is satisfied.
\end{proof}

Last lemma is directly derived from \cite{blavatskyy2010}'s Corollary 2. Since \cite{blavatskyy2010} starts from a general theorem which is not needed in this paper, axioms and Corollary 2 are simplified. The notation $p^N_i$ is applied here to represent the secured winning probability of contestant $i$, while that of a draw among all contestants is denoted by $p_\mathrm{null}^N: \mathbb{R}_+^N \rightarrow [0, 1]$ to represent the null case, where the prize is not secured to specific contestant.\footnote{In \cite{blavatskyy2010}, a series of more general notations are applied. Denote by $\Omega = 2^N\setminus\varnothing$ all possible non-empty subsets of contestant set $N$, where $2^N$ is the power set. For any group of contestants $S\in\Omega$, the notation $p^S(x)$ is used to represent the probability of contest ending up with a draw among contestants $S$ while other contestants $N\setminus S$ lose. For example, $p^{\{i\}}$ means the probability that contestant $i$ wins without a draw, $p^N(x)$ means the probability that a draw happens among all contestants. In this paper only corollary 2 of \cite{blavatskyy2010} about the draw with two level (either individual or all contestants) is used, where the possibilities of the draw between these two levels are restricted to zero, i.e. $p^S(x) = 0$ for all $S\in\Omega: 2\leq|S|\leq n-1$. So the notation is simplified and adapted consistently in this paper. $p^{\{i\}}(x)$ is simplified as $p^N_i(x)$ to represent the probability that contestant $i$ secure the winning without a draw, and $p^N(x)$ is transferred to $p_\mathrm{null}^N(x)$ to represent the probability that all contestants end up with a draw. Since the draw is not defined exactly for specific contestant. It is interpreted as null case in this paper to capture all events out of individual success.} A partial allocation model with positive probability of the null case $p^N_\mathrm{null}(x) = 1 - \sum_{i\in N} p^N_i(x) > 0$ is considered.\footnote{Note that in \cite{blavatskyy2010}, SM is slightly different from that in this paper. So that positive probability of the draw must hold, which is assumed in the partial allocation model. Here we set partial allocation as given and restrict the domain into $X$.  Readers can check that the same functional form still holds.} An additional axiom is introduced and \cite{blavatskyy2010}'s Corollary 2 follows. 

\noindent\textbf{Draw independence (DI)}: For all $i\in N$ and all $x\in X$, there exists a function $g_i:\mathbb{R}_+\rightarrow[0, 1]$ such that 
\begin{equation*}
    \frac{p^N_i(x)}{1 - \sum_{j\in N\setminus \{i\}} p^N_j(x)} = g_i(x_i).
\end{equation*}

\begin{lemma}[\citealp{blavatskyy2010}] \label{lemma:blavatskyy}
    In a partial allocation model, SM, DI and p-RH hold if and only if for all $i\in N$,
    \begin{equation}
        p^N_i(x) = \frac{a_i x_i^r}{1 + \sum_{j\in N} a_j x_j^r}, \label{eq:blav_ind}
    \end{equation}
    for some $r> 0$ and $a_j > 0$ for all $j\in N$. $p^N_\mathrm{null}(x)$ is derived accordingly.
\end{lemma}

\subsubsection{Main proof}\label{apx:ind1.2}
(only if) The proof follows 5 steps proposed in Section \ref{sec:ind1}. 

Step 1: From Lemma \ref{lemma:logit}, SM and LCA imply an asymmetric logit CSF as in (\ref{eq:logit}). The deviation in contestant $j$'s probabilistic allocation from contestant $i$ being active can thus be derived from (\ref{eq:logit}) in the form of $f_j(\cdot)$: 
\begin{equation}
    d_{ij}(x) 
    = \frac{p^N_j(0, x_{-i})-p^N_j(x)}{p^N_j(0, x_{-i})} 
    = \frac{\frac{f_j(x_j)}{f_i(0) + \sum_{k\in N\setminus \{i\}}f_k(x_k)} - \frac{f_j(x_j)}{\sum_{k\in N}f_k(x_k)}}{\frac{f_j(x_j)}{f_i(0) + \sum_{k\in N\setminus \{i\}}f_k(x_k)}} 
    = \frac{f_i(x_i) - f_i(0)}{\sum_{k\in N} f_k(x_k)} \label{eq:deviation}
\end{equation}

Step 2: Without loss of generality, suppose contestant 1 always be active. Constructing new individual CSFs and null CSF as follows:
\begin{equation}
    \mu^N_i(x) = p^N_i(x) - p^N_i(0, x_{-i})\frac{p^N_1(x)}{p^N_1(0, x_{-i})}, \label{eq:ind12}
\end{equation}
for all $i\in N\setminus\{1\}$. Choose another contestant $j$ with positive allocation $p^N_j(x)>0$.
Individual CSF of contestant 1 is constructed as
\begin{equation}
    \mu^N_1(x) = p^N_1(x) - p^N_1(0, x_{-1})\frac{p^N_j(x)}{p^N_j(0, x_{-1})}.\label{eq:ind13}
\end{equation}
Substituting (\ref{eq:logit}) into (\ref{eq:ind12}) and (\ref{eq:ind13}), there is
\begin{equation}
    \mu^N_i(x) 
    = \frac{f_i(x_i)}{\sum_{k\in N} f_k(x_k)} - \frac{f_i(0)}{f_i(0) + \sum_{k\in N\setminus\{i\}} f_k(x_k)} \frac{\frac{f_1(x_1)}{\sum_{k\in N} f_k(x_k)}}{\frac{f_1(x_1)}{f_i(0) + \sum_{k\in N\setminus\{i\}} f_k(x_k)}} = \frac{f_i(x_i) - f_i(0)}{\sum_{k\in N} f_k(x_k)}, \label{eq:mu_ind}
\end{equation}
for all $i\in N$. Notice that new constructed individual CSF $\mu^N_i(x)$ is identical to the deviation (\ref{eq:deviation}). Thus d-RH of asymmetric logit CSF (\ref{eq:logit}) implies p-RH of newly constructed individual CSF (\ref{eq:mu_ind}). Also, we construct the null CSF according to (\ref{eq:mu_ind}). That is,
\begin{equation}
    \mu^N_{\mathrm{null}}(x) = 1 - \sum_{i\in N}\mu^N_i(x) = \frac{\sum_{i\in N} f_i(0)}{\sum_{k\in N} f_k(x_k)}. \label{eq:mu_null}
\end{equation}

Step 3: Consider the case of $\sum_{i\in N} f_i(0) = 0$, which implies $f_i(0) = 0$ for all $i\in N$. Notice that in this case, $\mu^N_\mathrm{null}(x) = 0$ and $\mu^N_i(x)$ is reduced into $p^N_i(x)$ as derived before. From Lemma \ref{lemma:tullock}, the asymmetric Tullock CSF is derived as in (\ref{eq:ind11.1}). 

Step 4: Considering the case of $\sum_{i\in N} f_i(0) > 0$, where $\mu^N_\mathrm{null}(x) > 0$ always holds. Thus it is a partial allocation model. Since $f_i(0)$ is non-negative constant for all $i\in N$, strictly increasing $f_i(\cdot)$ implies the SM of constructed individual CSF. Notice also 
\begin{equation*}
    \frac{\mu^N_i(x)}{1 - \sum_{j\in N\setminus\{i\}} \mu^N_j(x)} 
    = \frac{\mu^N_i(x)}{\mu^N_i(x) + \mu_\mathrm{null}^N(x)} 
    = \frac{\frac{f_i(x_i) - f_i(0)}{\sum_{k\in N} f_k(x_k)}}{\frac{f_i(x_i) - f_i(0)}{\sum_{k\in N} f_k(x_k)} + \frac{\sum_{j\in N}f_j(0)}{\sum_{k\in N} f_k(x_k)}} 
    = \frac{f_i(x_i) - f_i(0)}{f_i(x_i) + \sum_{j\in N\setminus\{i\}} f_j(0)},
\end{equation*}
which is a function of $x_i$ only and the range is covered by $[0, 1]$.
Accompany with p-RH being checked for new constructed individual CSFs in Step 2, Lemma \ref{lemma:blavatskyy} shows that individual CSF is in the form of (\ref{eq:blav_ind}) and the null CSF is defined accordingly. Combining them with (\ref{eq:mu_ind}) and (\ref{eq:mu_null}), respectively, there are
\begin{subequations}
\begin{align}
    \frac{f_i(x_i) - f_i(0)}{\sum_{k\in N} f_k(x_k)} 
    &= \frac{a_i x_i^r}{1 + \sum_{k\in N} a_k x_k^r}; \label{eq:mu_blav_ind} \\
    \frac{\sum_{k\in N} f_k(0)}{\sum_{k\in N} f_k(x_k)} 
    &= \frac{1}{1 + \sum_{k\in N} a_k x_k^r}. \label{eq:mu_blav_null}
\end{align}
\end{subequations}
From (\ref{eq:mu_blav_null}), there is
\begin{equation*}
    \sum_{k\in N} f_k(x_k) = \Bigg(1 + \sum_{k\in N}a_k x_k^r\Bigg) \sum_{k\in N} f_k(0).
\end{equation*}
Substituting it into (\ref{eq:mu_blav_ind}), it is
\begin{equation*}
    \frac{f_i(x_i) - f_i(0)}{(1 + \sum_{k\in N}a_k x_k^r) \sum_{k\in N} f_k(0)} = \frac{a_i x_i^r}{1 + \sum_{k\in N} a_k x_k^r},
\end{equation*}
which is equivalent to
\begin{equation*}
    f_i(x_i) = f_i(0) +\Bigg[a_i \sum_{k\in N} f_k(0)\Bigg] x_i^r.
\end{equation*}
Thus 
\begin{equation*}
    p^N_i(x) 
    = \frac{f_i(x_i)}{\sum_{j\in N} f_j(x_j)} = \frac{f_i(0) + [a_i \sum_{k\in N} f_k(0)] x_i^r}{\sum_{j\in N} \{f_j(0) + [a_j \sum_{k\in N} f_k(0)] x_j^r\}} 
    = \frac{\frac{f_i(0)}{\sum_{k\in N} f_k(0)} + a_i x_i^r}{\sum_{j\in N} \Big[\frac{f_j(0)}{\sum_{k\in N} f_k(0)} + a_j x_j^r\Big]}.
\end{equation*}
Let $b_j = f_j(0)/\sum_{k\in N} f_k(0)$. It is
\begin{equation}
    p^N_i(x) = \frac{b_i + a_i x_i^r}{\sum_{j\in N} (b_j + a_j x_j^r)},  \label{eq:main}
\end{equation}
for all $i\in N$. Notice that (\ref{eq:main}) holds only for $\sum_{i\in N} b_i\in (0, 1]$ so far. Multiplying parameters $a_j, b_j$ for all $j\in N$ by a positive constant leads to an unbounded parameter space, which allows $\sum_{i\in N} b_i > 0$.
    
Step 5: Combining CSF forms derived from both cases in former two steps, we know that (\ref{eq:main}) holds for $\sum_{i\in N} b_i \geq 0$. Sub-contest representations (\ref{eq:mainluck}) can be derived from LCA.

(if) The CSF (\ref{eq:mainluck}) is a special case of asymmetric logit CSF (\ref{eq:ind11}). Thus SM and LCA are satisfied. Now check d-RH. The deviation is
\begin{equation*}
    d_{ij}(x) 
    = \frac{p^N_j(0, x_{-i})-p^N_j(x)}{p^N_j(0, x_{-i})}
    = \frac{\frac{b_j + a_j x_j^r}{b_i + \sum_{k\in N\setminus\{i\}} (b_k + a_k x_k^r)} - \frac{b_j + a_j x_j^r}{\sum_{k\in N} (b_k + a_k x_k^r)}}{\frac{b_j + a_j x_j^r}{b_i + \sum_{k\in N\setminus\{i\}} (b_k + a_k x_k^r)}} 
    = \frac{a_i x_i^r}{\sum_{k\in N} (b_k + a_k x_k^r)}. 
\end{equation*}
So that
\begin{equation*}
    \frac{d_{ij}(x)}{d_{ji}(x)} 
    = \frac{\frac{a_i x_i^r}{\sum_{k\in N} (b_k + a_k x_k^r)}}{\frac{a_j x_j^r}{\sum_{k\in N} (b_k + a_k x_k^r)}} 
    = \frac{a_i x_i^r}{a_j x_j^r} = \frac{a_i (\lambda x_i)^r}{a_j (\lambda x_j)^r} 
    = \frac{\frac{a_i (\lambda x_i)^r}{\sum_{k\in N} [b_k + a_k (\lambda x_k)^r]}}{\frac{a_j (\lambda x_j)^r}{\sum_{k\in N} [b_k + a_k (\lambda x_k)^r]}} 
    = \frac{d_{ij}(\lambda x)}{d_{ji}(\lambda x)}.
\end{equation*}
Thus d-RH is satisfied.
\qed

\subsubsection{Alternative proof} \label{apx:ind1.3}
Theorem \ref{thm:main} can also be proved directly without help of Lemmata \ref{lemma:tullock} and \ref{lemma:blavatskyy}. We replace Steps 2, 3, and 4 of (only if) part of the main proof by the following. 

d-RH implies that for all $i,j\in N, x_i, x_j > 0, \lambda > 0$,
\begin{equation}
    \frac{f_i(x_i) - f_i(0)}{f_j(x_j) - f_j(0)} = \frac{f_i(\lambda x_i) - f_i(0)}{f_j(\lambda x_j) - f_j(0)}.\label{eq:ind15} 
\end{equation}
Let $F_i(z) = f_i(z) - f_i(0)$ for all $i\in N$. From Step 1, $f_i(z)$ is a strictly increasing function and $f_i(0)\geq 0$. Thus $F_i(z)$ is strictly increasing and $F_i(0) = f_i(0) - f_i(0) = 0$. We rewrite (\ref{eq:ind15}) as
\begin{equation}
    \frac{F_i(\lambda x_i)}{F_i(x_i)} = \frac{F_j(\lambda x_j)}{F_j(x_j)}.\label{eq:alter1}
\end{equation}
Notice that the LHS of (\ref{eq:alter1}) is independent to $x_j$. So it applies to any positive real number. There is
\begin{equation*}
    \frac{F_i(\lambda z)}{F_i(z)} = \frac{F_j(\lambda x_j)}{F_j(x_j)} = \frac{F_i(\lambda)}{F_i(1)},
\end{equation*}
which is expressed as follows after organizing:
\begin{equation*}
    \frac{F_i(\lambda z)}{F_i(1)} = \frac{F_i(\lambda)}{F_i(1)}\frac{F_i(z)}{F_i(1)}.
\end{equation*}
Let $f_i(z) = F_i(z)/F_i(1)$. It is $f_i(\lambda z) = f_i(\lambda)f_i(z)$, which is a Cauchy equation. The only continuous functional form is $f_i(z) = z^{r_i}$ \citep{aczel1966}. Let constant $F_i(1) = f_i(1) - f_i(0) = a_i > 0$. There is $F_i(z) = f_i(z)F_i(1) = a_i z^{r_i}$. Substituting it into (\ref{eq:alter1}), it is
\begin{equation*}
    \frac{a_i (\lambda x_i)^{r_i}}{a_i x_i^{r_i}} = \frac{a_j (\lambda x_j)^{r_j}}{a_j x_j^{r_j}}.
\end{equation*}
It derives that $\lambda^{r_i} = \lambda^{r_j}$, which holds for any pair $i, j\in N$. So that $r_i = r_j = r > 0$ for all $i, j\in N$, and thus $F_i(z) = a_i z^r$, where $r, a_i > 0$ for all $i\in N$. Let $f_i(0) = b_i\geq 0$. There is $f_i(z) = f_i(0) + F_i(z) = b_i + a_i z^r$, which holds for all $i\in N$. Thus the CSF (\ref{eq:mainluck}) is derived by substituting $f_j(x_j)$ for all $j\in N$ into (\ref{eq:ind11}). 
\qed

\subsubsection{Independence of axioms}\label{apx:ind1.4}
Now check that the axiomatization provided by the main theorem is tight. If SM is dropped, then the discriminating parameter $r$ can be negative. If d-RH is dropped, then the functional form will be exactly the one axiomatized by Lemma \ref{lemma:logit}. Following CSF illustrates a situation when LCA is dropped. For all $i\in M\subseteq N$, 
\begin{equation}
    p^M_i(x^M) = \frac{e^{\frac{x_i}{\sum_{k\in M} x_k}}}{\sum_{j\in M} e^{\frac{x_j}{\sum_{k\in M} x_k}}}.\label{eq:ind21}
\end{equation}
It's obvious that this CSF doesn't belong to the class of CSFs axiomatized by the main theorem. Indeed, it doesn't even belong to the asymmetric logit form (\ref{eq:ind11}). For any $i\in M\subset N$ with effort vector $(0, \mathbf{1}^{N\setminus\{i\}})$ and cardinality $m<n$, there is 
\begin{equation*}
\begin{split}
    p^N_i(0, \mathbf{1}^{N\setminus \{i\}}) 
    &= \frac{1}{1+(n-1)e^\frac{1}{n-1}} \\
    &\neq \frac{1}{1 + (m-1)e^\frac{1}{m-1}} \frac{1 + (m-1)e^\frac{1}{n-1}}{1+(n-1)e^\frac{1}{n-1}} \\
    &= p^M_i(0, \mathbf{1}^{M\setminus\{i\}})p^N_M(0, \mathbf{1}^{N\setminus \{i\}}),
\end{split}
\end{equation*}
which violates LCA.
At the same time, it satisfies SM, HOM, p-RH and d-RH. Thus LCA is required for the result.

Note that the model considered in this paper is based on an assumption $n\geq 3$. In the case of $n=2$, LCA becomes trivial. Thus the CSF (\ref{eq:ind21}), which satisfies SM and d-RH, proposes a counterexample for the result under $n=2$. The same form as shown in the main theorem for two contestants can only be obtained from sub-contest representation, which requires a larger full contest.

\subsection{Proof of Theorem \ref{thm:headonly}} \label{appendix:headonly}
(only if) From Lemma \ref{lemma:logit}, the asymmetric logit CSF (\ref{eq:ind11}) is derived. Since NAR applies for all $x$, fix a feasible $x$ such that $x_i + x_j = z > 0$. Let $x_i', x_j' \geq 0$ be redistributed efforts between $i,j$ such that $x_i + x_j = x_i' + x_j'$. Taking arbitrarily another contestant $k$, the probabilistic allocation of $k$ would be the same before and after the redistribution. That is,
\begin{equation*}
    \frac{f_k(x_k)}{f_i(x_i) + f_j(x_j) + \sum_{l\in N\setminus\{i,j\}}f_l(x_l)} = \frac{f_k(x_k)}{f_i(x_i') + f_j(x_j') + \sum_{l\in N\setminus\{i,j\}}f_l(x_l)},
\end{equation*}
which is equivalent to 
\begin{equation}
    f_i(x_i) + f_j(x_j) = f_i(x_i') + f_j(x_j'). \label{eq:headonly1}
\end{equation}
Let $x_i = x_j' = z$ and $x_i' = x_j = 0$. It becomes
\begin{equation*}
    f_i(z) - f_i(0) = f_j(z) - f_j(0),
\end{equation*}
which holds for all $i, j\in N$. Define a new function $F(z) = f_i(z) - f_i(0)$ for all $i\in N$. Subtracting $f_i(0) + f_j(0)$ on both sides of (\ref{eq:headonly1}), there is
\begin{equation*}
    [f_i(x_i) - f_i(0)] + [f_j(x_j) - f_j(0)] = [f_i(x_i') - f_i(0)] + [f_j(x_j') - f_j(0)],
\end{equation*}
which is equivalent to
\begin{equation}
    F(x_i) + F(x_j) = F(x_i') + F(x_j').\label{eq:ind16}
\end{equation}
Noticed also $F(0) = f_i(0) - f_i(0) = 0$. Let $x_i = t$, $x_j = s$, $x_j' = 0$, and $x_i' = t + s$. (\ref{eq:ind16}) implies that
\begin{equation*}
    F(t) + F(s) = F(t + s),
\end{equation*}
which is Cauchy's basic equation. Thus the form should be $F(z) = az$ for some constant $a$ \citep{aczel1966}. Strictly increasing $f_i(\cdot)$ implies that $F(z)$ is a strictly increasing function. Thus $a>0$. Let $f_j(0) = a b_j$ for all $j\in N$. There is $f_j(z) = F(z) + f_j(0) = a(b_j + z)$. Substituting it back to (\ref{eq:ind11}), it derives the functional form (\ref{eq:ind11.4}).

(if) The CSF (\ref{eq:ind11.4}) is a special case of asymmetric logit CSF (\ref{eq:ind11}). Thus SM and LCA are satisfied. Now check NAR. If $x_i + x_j = 0$, other contestants' probabilistic allocation will remain the same for sure. Arbitrarily choosing contestants $i,j\in N$ with $x_i + x_j = x_i' + x_j'> 0$. Taking another contestant $k\in N\setminus\{i,j\}$, there is
\begin{equation*}
\begin{split}
    p^N_k(x) 
    &= \frac{b_k + x_k}{b_i + x_i + b_j + x_j + \sum_{l\in N\setminus\{i,j\}}(b_l + x_l')} \\
    &= \frac{b_k + x_k}{b_i + x_i' + b_j + x_j' + \sum_{l\in N\setminus\{i,j\}}(b_l + x_l')} \\
    &= p^N_k(x_i', x_j', x^{N\setminus\{i,j\}}).
\end{split}
\end{equation*}
Thus NAR is satisfied.
\qed

\subsection{Proofs of Propositions \ref{prop:dummy} and \ref{prop:dummy2}}
\subsubsection{Proof of Proposition \ref{prop:dummy}} \label{appendix:dummy}
CRI gives that $p^N_i(0, x_{-j}) = p^N_i(x)/[1-p^N_j(x)]$. DC says that $p^N_i(0, x_{-j}) = p_i^{N\setminus\{j\}}(x^{N\setminus\{j\}})$. Combining with the fact that $\sum_{k\in N\setminus\{j\}}p_k^N(x) = 1-p^N_j(x)$, LCA implies $p_i^{N\setminus\{j\}}(x^{N\setminus\{j\}}) = p^N_i(x)/[1-p^N_j(x)]$. Taking LCA as given, a series of equality can be derived from either CRI or DC. That is,
\begin{equation*}
    p^N_i(0, x_{-j}) = \frac{p^N_i(x)}{1-p^N_j(x)} = p_i^{N\setminus\{j\}}(x^{N\setminus\{j\}}).
\end{equation*}
Hence it proves the equivalence.
\qed

\subsubsection{Proof of Proposition \ref{prop:dummy2}} \label{apx:ind3.2}
Following lemma is proposed first to support the proof of Proposition \ref{prop:dummy2}. 
\begin{lemma} \label{lem:ind5}
    SM and LCA imply that for all $x\in X$, all $i\in N$, all $x_i'>0$, and all $j\in N\setminus\{i\}$,
    \begin{equation*}
        x_i < x_i'\quad \Rightarrow \quad p^N_j(x) \geq p^N_j(x_i', x_{-i}),
    \end{equation*} 
    where equality only holds when $x_j = 0$ and $p^N_j(x) = 0$.
\end{lemma}

\begin{proof}
From Lemma \ref{lemma:logit}, SM and LCA imply an asymmetric logit form (\ref{eq:logit}) with strictly increasing $f_j(\cdot)\geq 0$. Suppose $x_i < x_i'$. There is $f_i(x_i) < f_i(x_i')$.
Notice that the CSF for any other contestant $j\in N\setminus\{i\}$ has only $f_i(\cdot)$ in denominator. Thus 
\begin{equation*}
    p^N_j(x_i, x_{-i}) = \frac{f_j(x_j)}{f_i(x_i) + \sum_{k\in N\setminus\{i\}} f_k(x_k)} \geq \frac{f_j(x_j)}{f_i(x_i') + \sum_{k\in N\setminus\{i\}} f_k(x_k)} = p^N_j(x_i', x_{-i}),
\end{equation*}
where equality obtains at $f_j(x_j) = 0$.
\end{proof}

This lemma says that contestant's allocation is decreasing in efforts of other contestants. Usually it is included in SM for many papers in literature, e.g., \cite{skaperdas1996, clark1998}. This lemma claims that it can also be derived from existing axioms. 


\begin{proof}[Proof of Proposition \ref{prop:dummy2}]
Suppose $x$ is an effort vector of contestants $N$ with at least one active contestant and one inactive contestant. Without loss of generality, suppose contestant 1 is active and contestant $n$ is inactive and there exists an integer $t^*\leq n-1$ such that $x_i > 0$ for all $i \leq t^*$, and $x_i = 0$ for all $i\geq t^* + 1$. 

Case 1: $t^* = 1$. Suppose that $p^N_1(x) < 1$. Taking $\lambda > 1$, there is $x_1 < \lambda x_1$. Since all other contestants invest zero, thus $\lambda x = (\lambda x_1, x_2, \cdots, x_n)$. SM implies that $p^N_1(\lambda x) > p^N_1(x)$ in terms of $p^N_1(x)< 1$, which violates HOM. Thus $p^N_1(x) = 1$ must hold by contradiction. It also implies that $\sum_{i\in N\setminus\{1\}} p^N_i(x) = 0$ and thus $p^N_i(x) = 0$ for all $i\in N\setminus\{1\}$. Arbitrarily choose a dummy $i\in N\setminus\{1\}$. From LCA and $p^N_i(x) = 0$, there is 
\begin{equation*}
    p_j^{N\setminus\{i\}}(x^{N\setminus\{i\}}) = \frac{p^N_j(x)}{\sum_{k\in N\setminus\{i\}}p_k^N(x)} = \frac{p^N_j(x)}{1 - p^N_i(x)} = p^N_j(x),
\end{equation*} 
for all $j\in N\setminus\{i\}$. Since $i, j$ are arbitrarily chosen, DC is satisfied in this case.

Case 2: $t^* \geq 2$. Taking $\lambda > 1$, sequentially change $x_i$ to $\lambda x_i$ from $i = 1$ to $t^*$ and thus effort vector $x$ would be transferred into $\lambda x$. From Lemma \ref{lem:ind5}, the allocations for all contestants $j\geq t^* + 1$ satisfy $p^N_j(x) \geq p^N_j(\lambda x)$ for all $j\geq t^* + 1$, where the equality holds only for $p^N_j(x) = 0$. From HOM, there is $p^N_i(x) = p^N_i(\lambda x)$ for all $i\leq t^*$, and thus $\sum_{i=1}^{t^*} p^N_i(x) = \sum_{i=1}^{t^*} p^N_i(\lambda x)$. Since the summation is 1, it also implies that $\sum_{j=t^*+1}^{n} p^N_j(x) = \sum_{j=t^*+1}^{n} p^N_j(\lambda x)$, which can only hold for $p^N_j(x) = 0$ for all $j\geq t^*+1$. Arbitrarily choosing one dummy $j\geq t^*+1$. From LCA and $p^N_j(x) = 0$, there is 
\begin{equation*}
    p_i^{N\setminus\{j\}}(x^{N\setminus\{j\}}) = \frac{p^N_i(x)}{\sum_{k\in N\setminus\{j\}}p_k^N(x)} = \frac{p^N_i(x)}{1 - p^N_j(x)} = p^N_i(x),
\end{equation*}
for all $i\in N\setminus\{j\}$. Therefore DC is implied.
\end{proof}

\subsection{Proof of Theorem \ref{thm:ind4}}\label{apx:ind8}
(only if) From IIE, we can easily obtain that 
for all $I\subset N$, $g_i^I(x^I)\geq 0$ for all $i\in I$, and $\sum_{i\in I}g_i^I(x^I) \leq 1$ by summing up the equations repeatedly. 

Without loss of generality, denote contestant 1 the one who is always active and fix contestant 1's effort $\overline{x}_1$. From SM, we know that $p^N_1(x) > 0$ due to $\overline{x}_1 > 0$. Let $I = \{1\}$. IIE implies that there exists a function $g_1^{\{1\}}(\overline{x}_1)$ such that $p^N_1(x) = g_1^{\{1\}}(\overline{x}_1)[p^N_1(x) + p^N_{\mathrm{null}}(x)]$. Since $p^N_1(x) > 0$, there is $g_1^{\{1\}}(\overline{x}_1) > 0$. So that 
\begin{equation}
    p^N_{\mathrm{null}}(x) = \frac{1-g_1^{\{1\}}(\overline{x}_1)}{g_1^{\{1\}}(\overline{x}_1)} p^N_1(x).\label{eq:ind34}
\end{equation}
Note that $g_1^{\{1\}}(\overline{x}_1)$ is a constant, as $\overline{x}_1$ is fixed. Denote $z = [1-g_1^{\{1\}}(\overline{x}_1)]/g_1^{\{1\}}(\overline{x}_1) \geq 0$. Notice also that $p^N_{\mathrm{null}}(x)=0$ always holds when $g_1^{\{1\}}(\overline{x}_1)=1$. 

For all $i\in N\setminus\{1\}$, we denote $I = \{1, i\}$. IIE implies that there exists a function $g_i^{\{1, i\}}(x_i; \overline{x}_1)$ such that $p^N_i(x) = g_i^{\{1, i\}}(x_i; \overline{x}_1)[p^N_1(x) + p^N_i(x) + p^N_{\mathrm{null}}(x)]$. Since $p^N_1(x) > 0$, we have $g_i^{\{1, i\}}(x_i; \overline{x}_1)\in (0, 1)$. Thus we obtain
\begin{equation}
    p^N_i(x) = \frac{g_i^{\{1, i\}}(x_i; \overline{x}_1)}{g_1^{\{1\}}(\overline{x}_1) [1 - g_i^{\{1, i\}}(x_i; \overline{x}_1)] } p^N_1(x).\label{eq:ind35}
\end{equation}

Let $f_1(\overline{x}_1) = 1$ and $f_i(x_i) \equiv g_i^{\{1, i\}}(x_i; \overline{x}_1)/g_1^{\{1\}}(\overline{x}_1) [1 - g_i^{\{1, i\}}(x_i; \overline{x}_1)] \geq 0$ for all $i\in N\setminus\{1\}$. Combining (\ref{eq:ind34}) and (\ref{eq:ind35}) together, as well as $p^N_{\mathrm{null}}(x) + p^N_1(x) + \sum_{i\in N\setminus\{1\}} p^N_i(x) = 1$, there is
\begin{equation*}
    p^N_1(x) 
    = \frac{f_1(\overline{x}_1)}{p^N_{\mathrm{null}}(x) + p^N_1(x) + \sum_{i\in N\setminus\{1\}} p^N_i(x)} p^N_1(x) 
    = \frac{f_1(\overline{x}_1)}{z + \sum_{j\in N} f_j(x_j)}.
\end{equation*}
Thus for all $i\in N$,
\begin{equation}
    p^N_i(x) = \frac{f_i(x_i)}{z + \sum_{j\in N} f_j(x_j)}.
\end{equation}
Taking this form in hand, SM implies that each $f_j(\cdot)$ is strictly increasing.

(if) SM is satisfied due to strictly increasing function $f_j(\cdot)$ for all $j\in N$. For all $i\in I\subset N$, we can find a function
\begin{equation*}
    g_i^I(x^I) = \frac{f_i(x_i)}{z + \sum_{j\in I} f_j(x_j)}.
\end{equation*}
So that
\begin{equation*}
    g_i^I(x^I)[1 - p^N_{N\setminus I}(x)] = \frac{f_i(x_i)}{z + \sum_{j\in I} f_j(x_j)}\frac{z + \sum_{j\in I} f_j(x_j)}{z + \sum_{j\in N} f_j(x_j)}
    = \frac{f_i(x_i)}{z + \sum_{j\in N} f_j(x_j)}
    = p^N_i(x).
\end{equation*}
Thus IIE is satisfied.
\qed

\end{document}